\documentclass[aps,prb,twocolumn,floatfix,amsmath,amssymb,superscriptaddress]{revtex4}
\usepackage[final]{graphicx}
\usepackage{bm}
\usepackage{afterpage}
\usepackage{color}
\usepackage{comment}
\usepackage[colorlinks=true,urlcolor=blue,linkcolor=blue,citecolor=blue]{hyperref}

\begin{document}

\title{Unconventional ferromagnetism and spin-triplet superconductivity in the imbalanced kagome-lattice Hubbard model}

\author{Chenyue Wen$^*$}
\affiliation{School of Physics, Beihang University,
Beijing, 100191, China}

\author{Xingchuan Zhu\footnote{These authors contributed equally to this work} }
\affiliation{Interdisciplinary Center for Fundamental and Frontier Sciences, Nanjing University of Science and Technology, Jiangyin, Jiangsu 214443, P. R. China}

\author{Ning Hao}
\affiliation{Anhui Key Laboratory of Condensed Matter Physics at Extreme Conditions, High Magnetic Field Laboratory, HFIPS, Anhui,
Chinese Academy of Sciences, Hefei, 230031, China}

\author{Huaiming Guo}
\email{hmguo@buaa.edu.cn}
\affiliation{School of Physics, Beihang University,
Beijing, 100191, China}
\affiliation{Beijing Computational Science Research Center, Beijing 100084, China}

\author{Shiping Feng}
\affiliation{ Department of Physics,  Beijing Normal University, Beijing, 100875, China}

\begin{abstract}
Unconventional ferromagnetism and superconductivity in the imbalanced kagome-lattice Hubbard model are investigated by the mean-field theory and determinant quantum Monte Carlo method. Due to the asymmetric band structure of kagome lattice,  the spin-$z$ ferromagnetic order intrinsically exists in the system, which is first enhanced by the interaction, and then continuously destructed after reaching a maximum at a moderate interaction strength. In contrast, the $xy$-plane ferromagnetism develops only above a critical interaction, which is estimated to be $U_c/t=3.65 \pm 0.05$ by finite-size scaling. We further verify the nature of the above transverse magnetic transition, and demonstrate it belongs to the three-dimensional $XY$ universality class. Finally, we study the superconducting property, and reveal the possible superconducting state has a triplet $f$-wave pairing symmetry. Our results uncover the exotic quantum states induced by the interactions on kagome lattice, and provide important insights regarding the interplay between electronic correlations and geometry frustrations.
\end{abstract}


\maketitle
\section{Introduction}
The Hubbard model is the simplest of all models describing interacting fermions in a lattice\cite{PhysRevLett.10.159,kanamori1963electron,hubbard1963electron}. It exhibits rich phases and phase transitions, and has become a paradigm in the field of strongly correlated electron systems. Up to now, the Hubbard model has been gaining resurgence of interest on various geometries, which is largely due to its relation to experiments in quantum materials.
Determinant quantum Monte Carlo (DQMC), as an unbiased numerical method, has been proved to be a powerful approach in investigating the Hubbard model\cite{PhysRevD.24.2278,Hirsch1985,PhysRevB.40.506,PhysRevB.39.839}. The behavior of the Hubbard model at half filling on two-dimensional bipartite lattices has been pinned down by DQMC very quantitatively, such as: the development of long-range antiferromagnetic (AF) order at infinitesimal values of $U$ on square lattice\cite{varney2009}, the nature of quantum criticality in the semi-metal to AF insulator transition on honeycomb lattice\cite{honeycomb1,assaad2013,honeycomb2,honeycomb3,honeycomb4,meng2010quantum}, et al..

Motivated by recent experimental findings, there is increasing interest in Hubbard model on frustrated lattices, where exotic quantum phase is expected to emerge. In the triangular lattice Hubbard model, an intermediated phase between metallic behavior at low interaction strength and Mott insulating spin-ordered phase at strong interactions is identified, and strong evidence points to that it is a gapped chiral spin liquid\cite{triangle1,triangle2,triangle3}. Compared to the triangular lattice, the kagome geometry has a smaller coordination number, thus can generate stronger frustration. The research on the interplay between correlation and frustration on kagome lattice is further boosted by the recent experimental discovery of several families of kagome materials exhibiting rich physics\cite{ortiz2019,ortiz2020,jiang2021unconventional,Yin_2021,ortiz2021,tan2021charge,FENG2021,zhou2021origin,wu2021nature,lin2021complex}. Related theoretical studies on the kagome lattice Hubbard model have revealed that there are no tendencies toward magnetic ordering and intermediate phases also exist at half filling\cite{kagome1,kagome2,wen2022}.

The existing theoretical results in the frustrated lattices are largely obtained by the density matrix renormalization group method on cylinders, since the DQMC method is usually blocked by the infamous sign problem due to the lack of particle-hole symmetry, thus are restricted to rather high temperatures. In the context of higher-order topological Mott insulator on kagome lattice, it is proposed that the sign problem can be avoided by simply changing the sign of the hopping amplitude of one spin species\cite{higher-order-kagome,otsuka2021higher}. Hence the resulting sign-problem-free Hamiltonian becomes an ideal platform to investigate the interplay between geometry frustration and electron-electron correlations for DQMC simulations. Then it is natural to ask what interesting quantum phases and phase transitions will occur in this modified kagome-lattice Hubbard model.

In this manuscript, we investigate the hopping-sign imbalanced kagome-lattice Hubbard model by means of two complementary methods: the mean-field theory and large-scale DQMC simulations. Since the energy spectrum of itinerant electrons on kagome lattice is asymmetric, the densities of the two spin subsystems are imbalanced at half filling, and  a spin-$z$ ferromagnetic (FM) order intrinsically exists in the system. Both methods reveal that: 1), the FM order in the $z$-direction is first enhanced by the Hubbard interaction, and then decreases gradually after reaching its maximum strength; 2), the $xy$-plane FM order develops only above a critical interaction. We perform a finite-size scaling of the DQMC data, and determine the critical value to be $U_c/t=3.65\pm 0.05$, which is much larger than that predicted by the mean-field approximation. The transverse magnetic transition is further verified to belong to the three-dimensional (3D) $XY$ universality class. Finally, we study the superconducting properties using DQMC, and find that if there is a possible superconducting instability, it would have a triplet $f$-wave symmetry.

This paper is organized as follows. Section II introduces the model we will investigate, along with our computational methodology.
Section III presents the results from the mean-field theory. Section IV uses DQMC simulations to study the magnetic properties of the imbalanced kagome-lattice Hubbard model. Section V demonstrates the superconducting properties. Section VI is the conclusions. Finally, in the appendices, we show the results for the imbalanced triangle-, square-, honeycomb-lattice Hubbard models, and more DQMC data on the imbalanced kagome lattice.

\begin{figure}[htbp]
\centering \includegraphics[width=8cm]{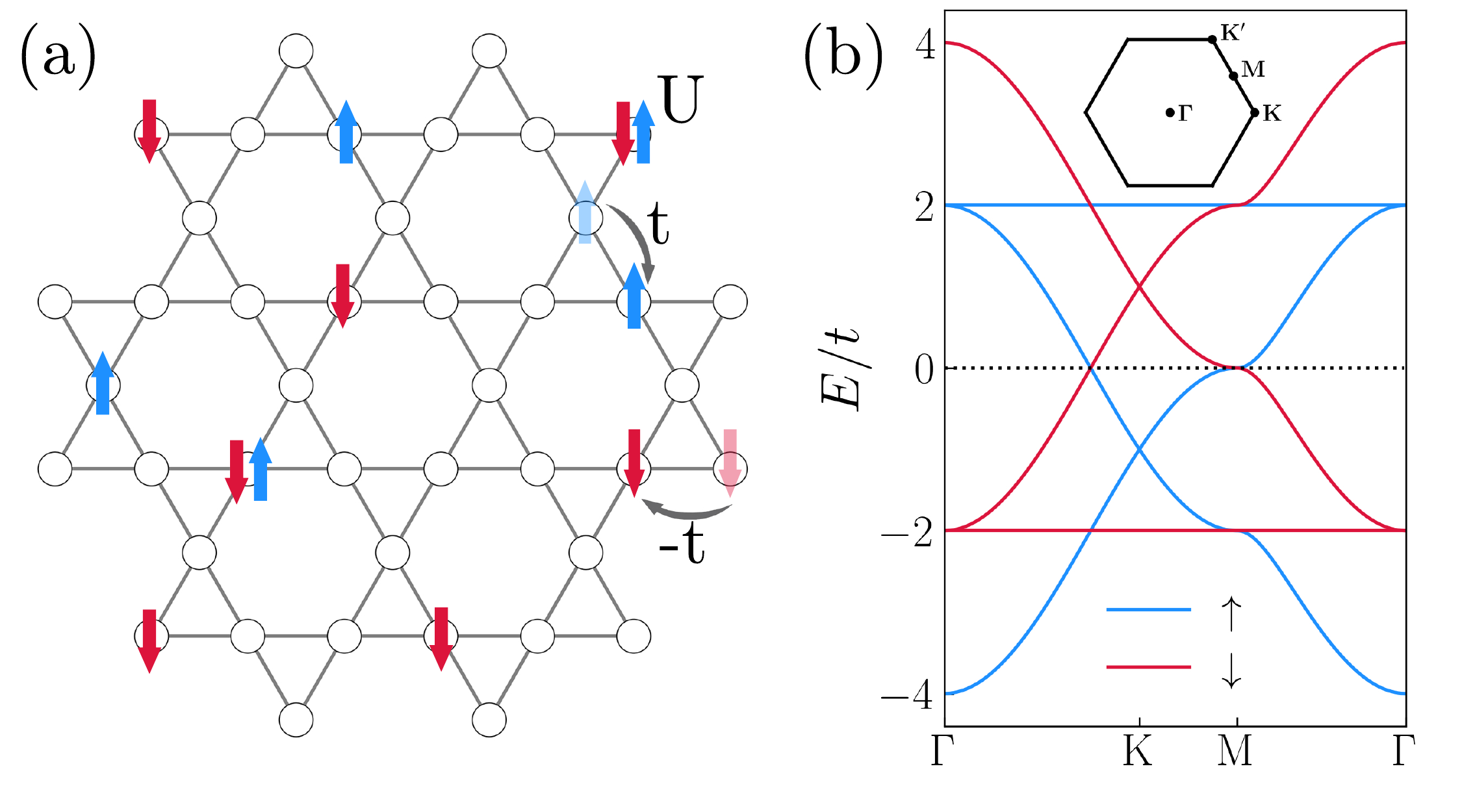} \caption{(a) A schematic show of the Hubbard model on kagome lattice, where $t$ is the hopping parameter and $U$ is the on-site Hubbard interaction. Up arrows and down arrows correspond to up-spin and down-spin electrons, respectively. (b) Band structure of the imbalanced tight-binding model [the non-interacting part in Eq.(1)] on kagome lattice. Inset in (b) is the first Brillouin zone, on which the high-symmetry points are marked.}
\label{fig1}
\end{figure}

\section{The model and method}

We start from the hopping-sign imbalanced kagome-lattice Hubbard model,

\begin{align}\label{eq1}
H=&-t\sum_{\langle ij\rangle}\sum_{\alpha,\beta=\uparrow,\downarrow}(c^{\dagger}_{i\alpha}\sigma^z_{\alpha\beta}c_{j\beta}+\textrm{H.c.}) \\ \nonumber
&+U\sum_{i}(n_{i\uparrow}-\frac{1}{2})(n_{i\downarrow}-\frac{1}{2})-\mu \sum_{i,\alpha}n_{i\alpha}
\end{align}
where $c_{i \alpha(\beta)}^{\dagger}$ and $c_{i \alpha(\beta)}$ are the creation and annihilation operators, respectively, at site $i$ with spin $\alpha(\beta)=\uparrow, \downarrow$; $\langle ij\rangle$ denotes nearest neighbors; $\sigma^z$ is the $z$-component of Pauli matrix, and results in opposite signs in the hopping amplitudes for spin-up and -down subsystems; $n_{i \alpha}=c_{i \alpha}^{\dagger} c_{i \alpha}$ is the number operator of electrons of spin $\alpha$ on site $i$; $U$ is the on-site repulsion, and $\mu$ is the chemical potential. We set the hopping amplitude $t = 1$ as the energy scale throughout the paper.

The kagome lattice has a three-site unit cell [Fig.~\ref{fig1}(a)]. In momentum space, the $U=0$ Hamiltonian is spin dependent, and is given by~\cite{guohm2009}
\begin{align}
\mathcal{H}_{0}^{\uparrow(\downarrow)}({\mathbf{k}})=\mp2 t\left(\begin{array}{ccc}
0 & \cos k_{1} & \cos k_{3} \\
\cos k_{1} & 0 & \cos k_{2} \\
\cos k_{3} & \cos k_{2} & 0
\end{array}\right),
\end{align}
where $k_n={\mathbf{k}}\cdot {\mathbf{a}}_n$ (the sublattice index $n=1,2,3$) with ${\mathbf{a}}_1=(1,0),{\mathbf{a}}_2=(-1,\sqrt{3})/2$, and ${\mathbf{a}}_3=-({\mathbf{a}}_1+{\mathbf{a}}_2)$. The whole spectrum has two flat bands $E^{\pm}_{3}({\mathbf{k}})=\pm 2t$, and four dispersive ones $
E^{\pm}_{1,2}({\mathbf{k}})=\pm t[-1\pm \sqrt{4f({\mathbf{k}})-3}]$,
with $f({\mathbf{k}})=\cos^2 k_1+\cos^2 k_2+\cos^2 k_3$.
For the dispersive bands, three momenta $M$ at the centers of the edges of the Brillouin zone (BZ) are saddle points, resulting in the Van Hove singularities (VHSs) at the filling $\rho=1/4$ and $5/12$, respectively. Although the spectrum of each individual Hamiltonian in Eq.(2) is asymmetric, the total one consisting of pairs of opposite bands is symmetric. As a consequence, $\mu/t=0$ corresponds to half filling, when the upper(lower) VHS of the spin-up(down) bands is exactly at the Fermi level. Besides, the spin-up and -down densities are imbalanced at half filling with $\rho_{\uparrow}<\rho_{\downarrow}$, which even persists for arbitrary $U$ and temperature $T$.

At finite interactions, Eq.\eqref{eq1} is solved numerically via DQMC, where one decouples the on-site interaction term through the introduction of an auxiliary Hubbard-Stratonovich field, which is integrated out stochastically. The only errors are those associated with the statistical sampling, the finite spatial lattice size, and the inverse temperature discretization. These errors are well controlled in the sense that they can be systematically reduced as needed, and further eliminated by appropriate extrapolations. The infamous sign problem generally exists for the non-bipartite kagome lattice~\cite{PhysRevB.41.9301,PhysRevLett.94.170201,PhysRevB.92.045110}. Yet it is accidentally eliminated in the spin-dependent Hamiltonian Eq.(\ref{eq1}), which can be demonstrated by a simple transformation $c_{i\uparrow}\rightarrow \tilde{c}_{i\uparrow}$ and $c_{i\downarrow}\rightarrow \tilde{c}_{i\downarrow}^{\dagger}$. It cancels the spin-dependent phase in the hopping, and changes the sign of the Hubbard interaction. The resulting Hamiltonian is just a normal attractive Hubbard model, thus is free of the sign problem at $\mu=0$.
Thus this specific model opens the door for DQMC to explore the exotic interacting quantum phases on highly frustrated kagome lattice. In the following, we use the inverse temperature discretization $\Delta\tau=0.1$, and the lowest temperature accessed is $T/t=1/25$. The lattice has $N=3\times L \times L$ sites with $L$ up to $12$.

\section{The mean-field theory}

\begin{figure}[htbp]
\centering \includegraphics[width=8.5cm]{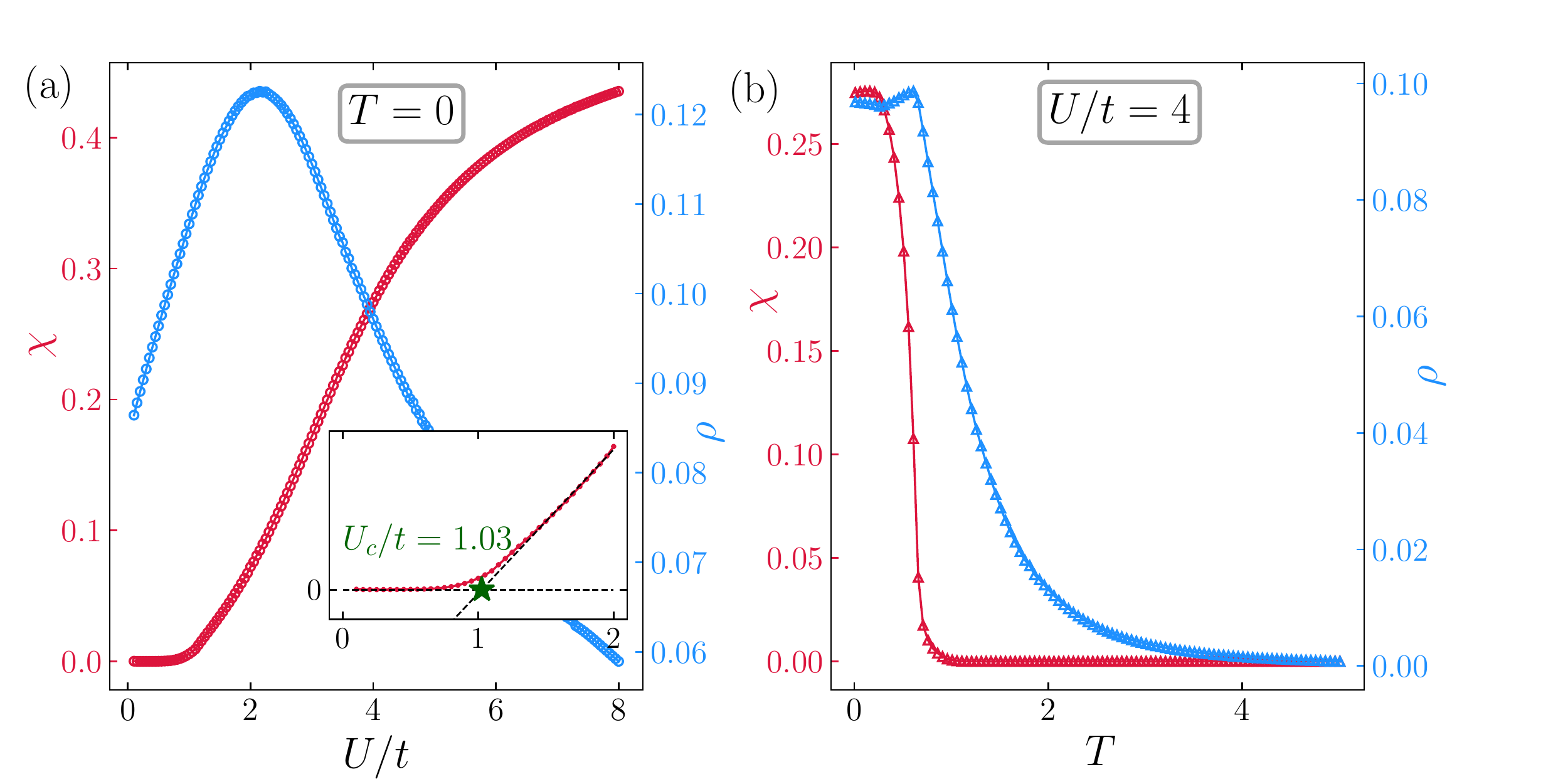} \caption{The order parameters $\chi,\rho$: (a) as a function of $U$ at $T=0$; (b) as a function of $T$ at $U/t=4$. Inset in (a) shows enlarged view of the transition region of $\chi$, and a linear fit of the curve gives a critical interaction $U_c/t=1.03$.}
\label{fig2}
\end{figure}

Within the mean-field approximation, the interacting term $Un_{i \uparrow} n_{i \downarrow}$ is decoupled as a Hartree term and a Fock term,
\begin{align}
n_{i \uparrow} n_{i \downarrow} \approx &n_{i \uparrow}\left\langle n_{i \downarrow}\right\rangle+\left\langle n_{i \uparrow}\right\rangle n_{i \downarrow}-\left\langle n_{i \uparrow}\right\rangle\left\langle n_{i \downarrow}\right\rangle \\ \nonumber
-&\left\langle S^{-}_i\right\rangle S^{+}_i-\left\langle S^{+}_i\right\rangle S^{-}_i+\left\langle S^{+}_i\right\rangle\left\langle S^{-}_i\right\rangle,
\end{align}
where $S^{+}_i=c_{i \uparrow}^{\dagger} c_{i \downarrow}$, and $S^{-}_i=c_{i \downarrow}^{\dagger} c_{i \uparrow}$.
The averages of the operators can be writen as $\langle n_{i \uparrow(\downarrow)}\rangle=\frac{1}{2}\mp \rho$ and $\langle c_{i \downarrow}^{\dagger} c_{i \uparrow}\rangle=\langle c_{i \uparrow}^{\dagger} c_{i \downarrow}\rangle=\chi$, with $\rho$ and $\chi$ the order parameters.
In the momentum space, the mean-field Hamiltonian is
\begin{align}
H_{MF}=\sum_{\bf k}\psi_{\bf k}^{\dagger}{\cal H}({\bf k})\psi_{\bf k}+E_0,\\ \nonumber
{\cal H}({\bf k})=\left(\begin{array}{cc}
H_{\uparrow \uparrow} & H_{\uparrow \downarrow}  \\
H_{\downarrow \uparrow} & H_{\downarrow \downarrow}  \\
\end{array}\right),
\end{align}
where $\psi_{\bf k}=(c_{1,{\bf k}},c_{2,{\bf k}},c_{3,{\bf k}})^{T}$ is a three-element basis; $H_{\uparrow \uparrow}=\mathcal{H}_{0}^{\uparrow}({\bf k})+U(\frac{1}{2}-\rho)$, $H_{\downarrow \downarrow}=\mathcal{H}_{0}^{\downarrow}({\bf k})+U(\frac{1}{2}+\rho)$, $H_{\uparrow \downarrow}=H_{\downarrow \uparrow}=-U\chi \mathbb{I}_{3\times 3}$, and the constant is $E_0=3NU\chi^2+3NU\rho^2-\frac{3}{4}NU$ with $N$ the total number of sites. Minimizing the free energy,
\begin{align}
F=-\frac{1}{\beta}\sum_{\bf k}\sum_{i=1}^{6}\ln (1+e^{-\beta E_{\bf k}^{(i)}})+E_0,
\end{align}
where $E_{\bf k}^{(i)}$ is the eigenenergy of the $i$th band at the momentum ${\bf k}$, the following self-consistent equations are obtained,
\begin{align}
\rho=-\frac{1}{6NU}\sum_{\bf k}\sum_{i=1}^{6}\frac{1 }{1+e^{\beta E_{\bf k}^{(i)}}}\frac{\partial E_{\bf k}^{(i)}}{\partial \rho}, \\ \nonumber
\chi=-\frac{1}{6NU}\sum_{\bf k}\sum_{i=1}^{6}\frac{1 }{1+e^{\beta E_{\bf k}^{(i)}}}\frac{\partial E_{\bf k}^{(i)}}{\partial \chi},
\end{align}
from which $\rho$ and $\chi$ can be calculated numerically.

Figure \ref{fig2}(a) shows the order parameters $\rho,\chi$ as a function of $U$ at zero temperature. Since the asymmetry of the bands between the two spin species leads to $\rho_{\uparrow}\neq \rho_{\downarrow}$ at $U=0$, a ferromagnetic order along the $z$ direction inherently exists in the system with the local magnetic moment $m=|\rho_{\uparrow}-\rho_{\downarrow}|=1/12$.
As $U$ increases, the value of $\rho$ increases accordingly until it reaches a maximum at a finite interaction, after which $\rho$ begins to decrease monotonically. In contrast, $\chi$ becomes nonzero only when $U$ is larger than a critical value $U_c$. By fitting the curve of $\chi$ linearly near the transition point, the critical interaction is determined to be $U_c/t=1.03$. For $U>U_c$, $\chi$ increases continuously and becomes saturated at extremely large $U$, implying the ferromagnetism is formed in the $xy$ plane, and gradually strengthened by the interaction. Since the modification of the hopping sign in the spin-down subsystem breaks the spin $SU(2)$ symmetry, the $z$-direction ferromagnetism becomes different from that in the $xy$ plane. Moreover, a finite temperature phase transition is expected to occur for both FM orders as $T$ is increased. Indeed, as shown in Fig.\ref{fig2}(b), $\chi,\rho$ gradually decrease with $T$, and become vanished at a critical temperature.

\section{DQMC study of the magnetic properties}

\begin{figure}[htbp]
\centering \includegraphics[width=8.5cm]{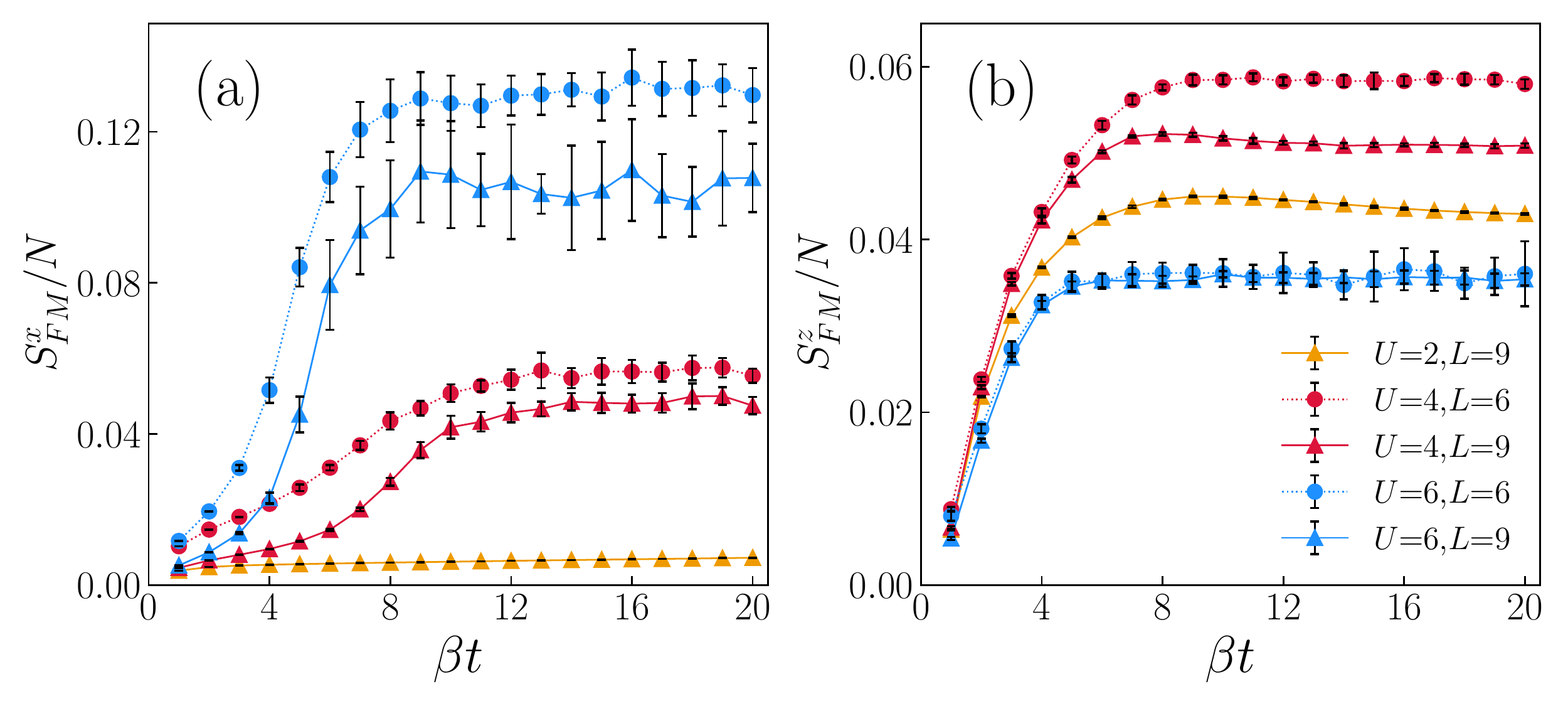} \caption{(a) $x$-component and (b) $z$-component of the FM structure factor as a function of inverse temperature for various $U$ and $L$.}
\label{fig3}
\end{figure}

\begin{figure}[htbp]
\centering \includegraphics[width=8.5cm]{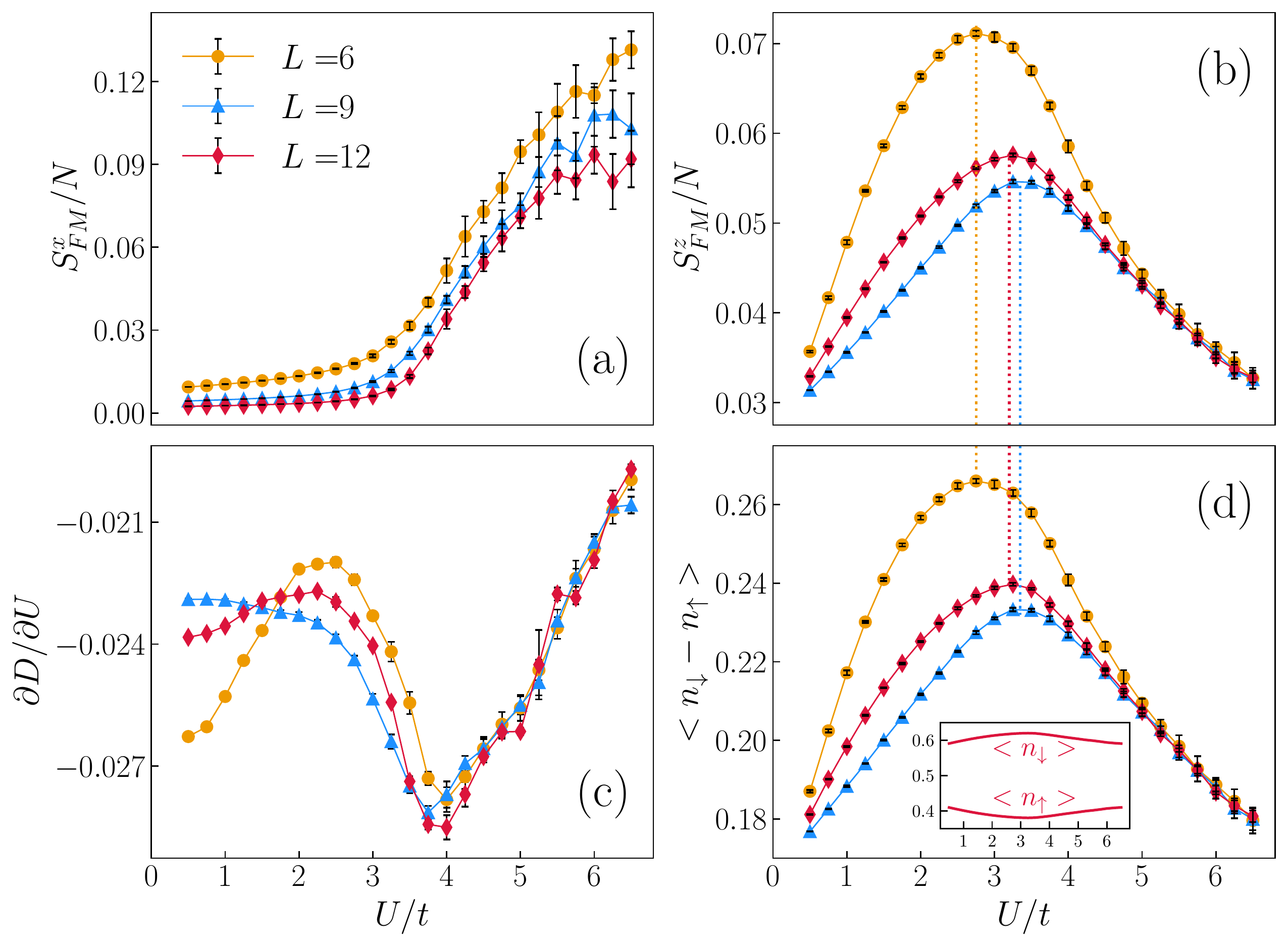} \caption{Several physical quantities as a function of $U$ for various $L$ at
$\beta t= 10$: (a) $x$-component and (b) $z$-component of the FM structure factor; (c) the rate of change of the double occupancy; (d) the spin-density imbalance in the $z$ direction. Inset in (d) plots the up- and down-spin average densites for the lattice size $L=12$. The dashed lines in (b) and (d) mark the peaks of the curves, which coincide with each other at the same lattice size.}
\label{fig4}
\end{figure}

Next we apply DQMC to unveil the physical behavior of the Hamiltonian in Eq.(1) quantitatively. The magnetic order is characterized by the static structure factor, which is defined by\cite{varney2009},
\begin{align}\label{eq1a}
  S^{\alpha}({\bf k})=\sum_{\bf l}e^{i{\bf k}\cdot{\bf l}}C^{\alpha}({\bf l}),
\end{align}
where $\alpha=x,z$ denotes the spin component, and the real-space spin-spin correlation function is defined as $C^{z}({\bf l})=\langle S^z_iS^z_{i+{\bf l}}\rangle$  and $C^{x}({\bf l})=\frac{1}{2}\langle S^x_iS^x_{i+{\bf l}}+S^y_iS^y_{i+{\bf l}}\rangle$. The ferromagnetism has an order vector ${\bf k}=0$, and we let $S_{FM}^{\alpha}=S^{\alpha}({\bf k}=0)$.

Figure \ref{fig3} shows the FM structure factor as a function of inverse temperature for various $L$ and $U$. $S^z_{FM}$ saturates to the ground-state value at large enough $\beta$. It is similar for $S^{x}_{FM}$ except for that the saturated value becomes finite only when $U$ is large enough. As $L$ increases, the average value of all spin-spin correlations, i.e., $S^{\alpha}/N$, decreases, and a larger value of $\beta$ is required to resemble the property of the ground state. As $U$ increases, $S^{x}_{FM}$ increases monotonically, but $S^z_{FM}$ does not. To see the FM evolution with $U$, we show $S^{\alpha}_{FM} (\alpha=x,y)$ versus the interaction strength for various $L$ at $\beta t=10$. As shown in Fig.\ref{fig4} (a), $S^{x}_{FM}$ remains negligibly small for weak interacting strength, and becomes finite for large $U$. This behavior indicates the FM order occurs at a critical interaction, which is consistent with the mean-field result. Along with this FM transition, there appears a peak in the absolute value of the rate of change of the double occupancy, which indicates $D=\langle n_{i\uparrow}n_{i\downarrow}\rangle$ is decreasing most rapidly here. The FM order parameter squared is estimated by finite size extrapolating $S^{x}_{FM}/N$ to $L\rightarrow \infty$. A non-vanishing value in the thermodynamic limit marks the formation of the FM order. Here the critical interaction is determined to be $U_c/t=3.65 \pm 0.05$.

\begin{figure}[htbp]
\centering \includegraphics[width=8.5cm]{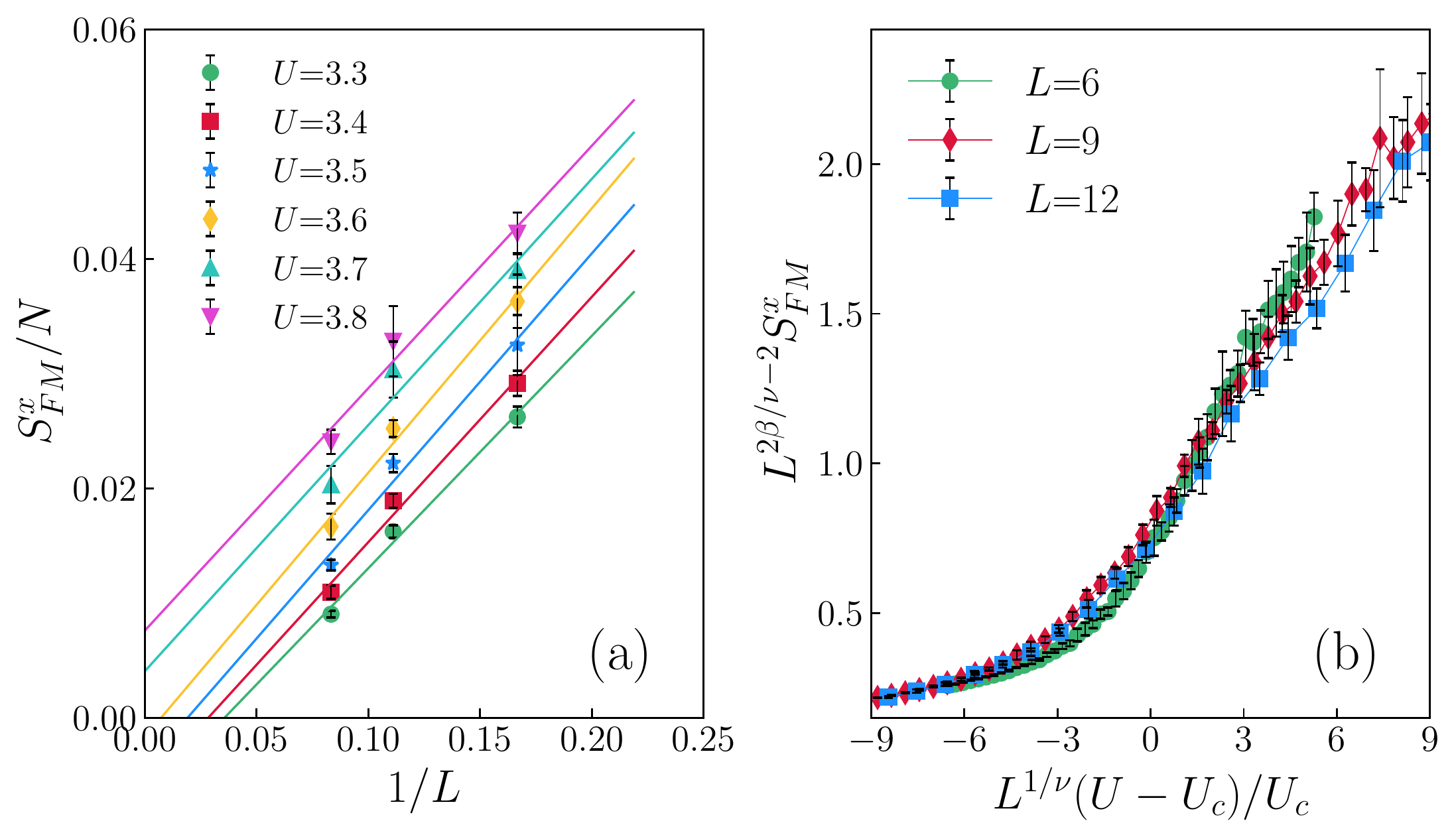} \caption{(a) Size dependence and extrapolation of the structure factor $S^x_{FM}/N$. The solid lines are least-squares fits to the linear form of $1/L$. The value in the thermodynamic limit becomes finite at some interaction between $U/t=3.6$ and $3.7$, thus the critical interaction is estimated to be $U_c/t=3.65\pm 0.05$. (b) The data collapse using the critical exponets of the three-dimensional $XY$ universality class and the critical interaction determined in (a).}
\label{fig5}
\end{figure}

We also perform a finite-size scaling analysis based on the usual scaling formula\cite{assaad2013},
\begin{align}
S_{FM}^{x}=L^{2-2\beta/\nu} F[L^{1/\nu}(U-U_c)],
\end{align}
where $\beta$ is the order parameter exponent, and $\nu$ is the correlation length exponent.
The FM transition in the $xy$ plane is expected to belong to the three-dimensional $XY$ universality class. The field-theoretical results obtained by analyzing the fixed-dimension expansion gives $\nu=0.6704(7)$ and the magnetic susceptibility exponent $\gamma=1.3164(8)$\cite{xyscale,vicari2002,zinn2021quantum}. Thus we have $2\beta=0.6948$ with the scaling relation $\gamma+2\beta=\nu d$.
Together with the above critical interaction determined by the finite-size scaling, we scale $S_{FM}^{x}$ at different lattice sizes according to Eq.(8). As shown in Fig.\ref{fig5}(b), the data collapse is quite good, and the nature of the phase transition is thus confirmed to belong to the three-dimensional $XY$ universality class.

Figure \ref{fig4}(b) plots $S^z_{FM}$ as a function of $U$, whose behavior is different from its counterpart in the $x$ direction. $S^z_{FM}$ first increases with $U$, and then reaches its peak value located at about $U_c$, after which it decreases gradually. The FM order should be closely related to the $z$-direction spin-density imbalance $ m_z=\langle n_{\uparrow}-n_{\downarrow}\rangle$. As shown in Fig.\ref{fig4}(d), $m_z$ changes with $U$ in exactly the same way with the static structure factor $S^z_{FM}$. At half filling, the local moment squared is $m^2=\langle (n_{\uparrow}-n_{\downarrow})^2\rangle=1-2n_{i\uparrow}n_{i\downarrow}$, which is negatively proportional to the double occupancy $D$. Our simulations find $D$ continuously decreases with $U$, which is expected since the on-site repulsion $U$ suppresses the double occupancy. Hence the local moment squared is continuously enhanced by the interaction. This can account for the initial increase of $S^z_{FM}$ at weak correlations.

In the large-$U$ limit, the double occupancy is completely eliminated, and the Hubbard model in Eq.(1) maps onto the following Heisenberg model\cite{charles1976},
\begin{align}
  {\cal H}=-J\sum_{\langle ij\rangle} (S^x_i S^x_j+S^y_i S^y_i)+J\sum_{\langle ij\rangle} S^z_iS^z_j,
\end{align}
where the exchange coupling is $J=\frac{4t^2}{U}$. The above spin Hamiltonian is unconventional in that it is ferromagnetic in the $xy$ plane, but antiferromagnetic in the $z$ direction. Generally there is a competition between the two kinds of magnetisms. Here the values of the different kinds of exchange couplings are the same. Since there are two components in the FM term, the model in Eq.(9) is expected to develop transverse long-range FM correlations rather than the longitudinal AF ones in the ground state.

\begin{figure}[htbp]
\centering \includegraphics[width=7.5cm]{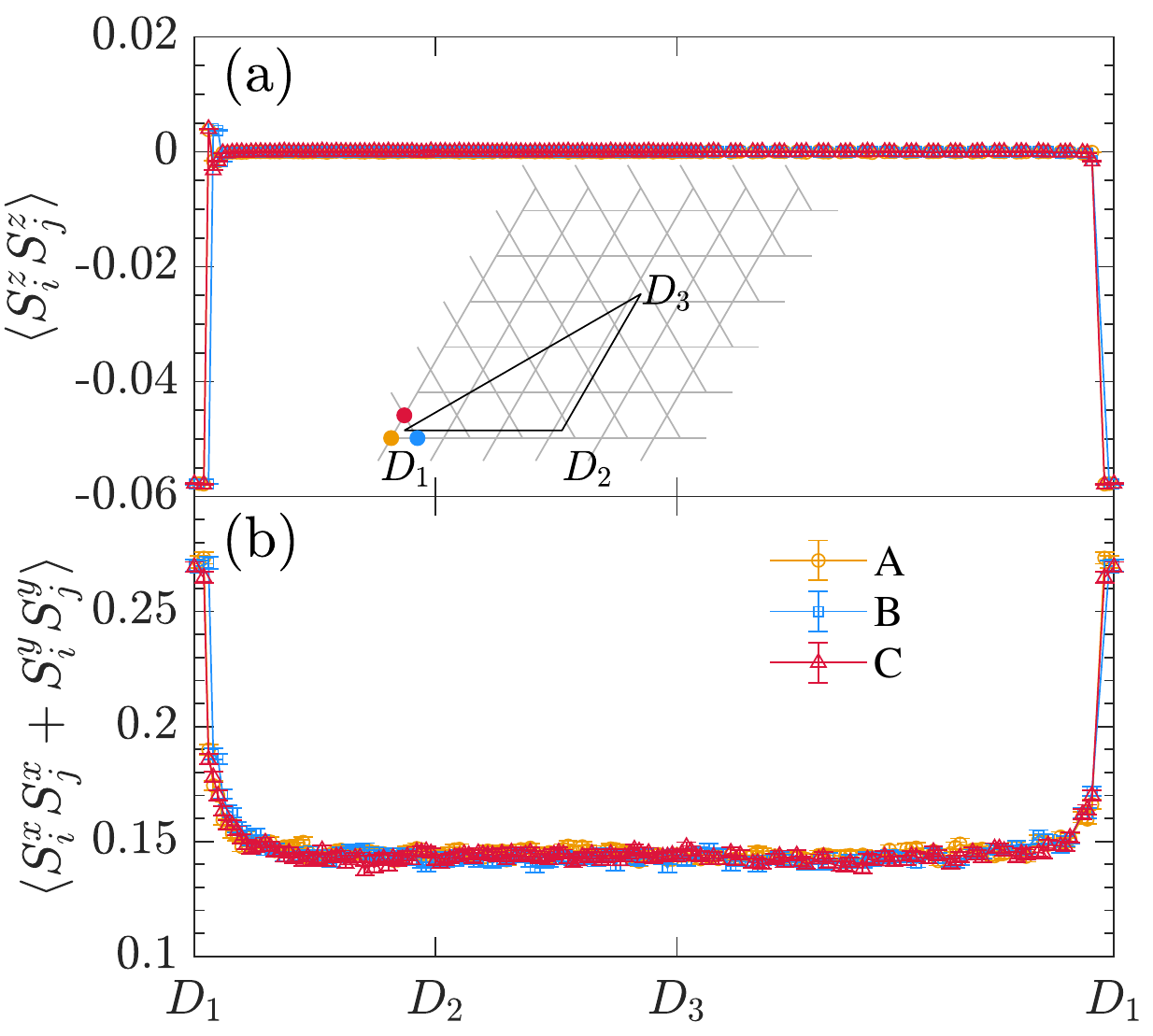} \caption{The longitudinal (a) and transverse (b) spin-spin correlations along the high-symmetry paths of the kagome geometry. The lattice size is $L=36$, and the inverse temperature is $\beta=2L$, which is low enough to ensure convergence into the ground state.}
\label{fig6}
\end{figure}

We have employed the approach of stochastic series expansion (SSE) QMC method with directed loop updates to study the above spin model Eq.(9)\cite{sandvik2002,syljuasen2003,fabien2005,pollet2004,Bauer2011}, and calculated the spin-spin correlations to characterize the long-range magnetic order. As shown in Fig.\ref{fig6}, the $z$ component of the spin correlation function $\langle S_i^zS_j^z\rangle$ decreases quickly as the sites $i,j$ depart from each other, and has already become zero at several-bond-length distance, implying there is no long-range order of $S_i^z$. In contrast, $\langle S_i^xS_j^x+S_i^yS_j^y\rangle$ decreases to an almost constant and finite value, which remains even for the largest distance in the lattice. Besides, the spin-spin correlations are all positive. Thus the ground state of the spin Hamiltonian Eq.(9) is ferromagnetically ordered, as expected. The spin-$\frac{1}{2}$ operators can be mapped to hardcore boson ones via $S_i^+=a_i^{\dagger}, S_i^-=a_i, S_i^z=a_i^{\dagger}a_i-1/2$, where $a^{\dagger}_i$ and $a_i$ are the hardcore-boson creation and annihilation operators, respectively\cite{10.1143/PTP.16.569}. The resulting hardcore Bose-Hubbard model on kagome lattice has been widely investigated in the literature\cite{Isakov2006,Isakov2006b,Isakov2007}. For the parameter values in Eq.(9), we have $V=2t, \mu=2V$, with $t$ the hopping amplitude, $V$ the nearest-neighbor interaction and $\mu$ the chemical potential in the mapped Bose-Hubbard model. At these parameters, hardcore bosons form superfluid, which is just the counterpart of the $xy$ FM order identified here.

\section{DQMC study of the superconducting properties}
It is well known that magnetism plays an important role in the emergence of superconductivity. A prominent example is cuprate superconductor, whose parent compound is an AF insulator. Superconductivity emerges from AF spin fluctuations in the doped system, and has unconventional spin-singlet pairing with $d$-wave symmetry. Here the system has long-range FM order at half filling, thus the novel spin-triplet pairing is highly expected to occur in a doped case. To explore the intriguing superconducting properties, we explore the uniform pairing susceptibility, which is defined as~\cite{PhysRevB.91.241107,PhysRevB.97.155146,PhysRevB.97.235453},
\begin{equation}
   \chi^{\alpha}=\frac{1}{N} \int_{0}^{\beta} d \tau \sum_{i j}\left\langle\Delta_{i}^{\alpha}(\tau) \Delta_{j}^{\alpha \dagger}(0)\right\rangle,
\end{equation}
where $ \Delta_{i}^{\alpha}(\tau)= \sum_{j} f_{i j}^{\alpha} e^{\tau H} \hat{P}_{ij}^{s} e^{-\tau H}$ is the time-dependent pairing operator with form-factor $f_{i j}^{\alpha}=0, \pm 1 \text { or } \pm 2$ for the bond
connecting sites $i$ and $j$, depending on the pairing symmetry $\alpha$. Since the spins are polarized in the FM state, we consider three kinds of pairings: $\hat{P}_{ij}^{1}=c_{i \uparrow} c_{j \uparrow}, \hat{P}_{ij}^{0}=c_{i \uparrow} c_{j \downarrow}, \hat{P}_{ij}^{-1}=c_{i \downarrow} c_{j \downarrow}$, corresponding to the total $z$-spin $s_z=1,0,-1$,respectively. The effective susceptibility, $\chi_{\text {eff }}^{\alpha}\equiv\chi^{\alpha}-\chi_{0}^{\alpha}$, subtracts the uncorrelated part $\chi_{0}^{\alpha}$ from $\chi^{\alpha}$, thereby directly capturing the interaction effects, and can be further used to evaluate the pairing vertex.

\begin{figure}[htbp]
\centering \includegraphics[width=8.5cm]{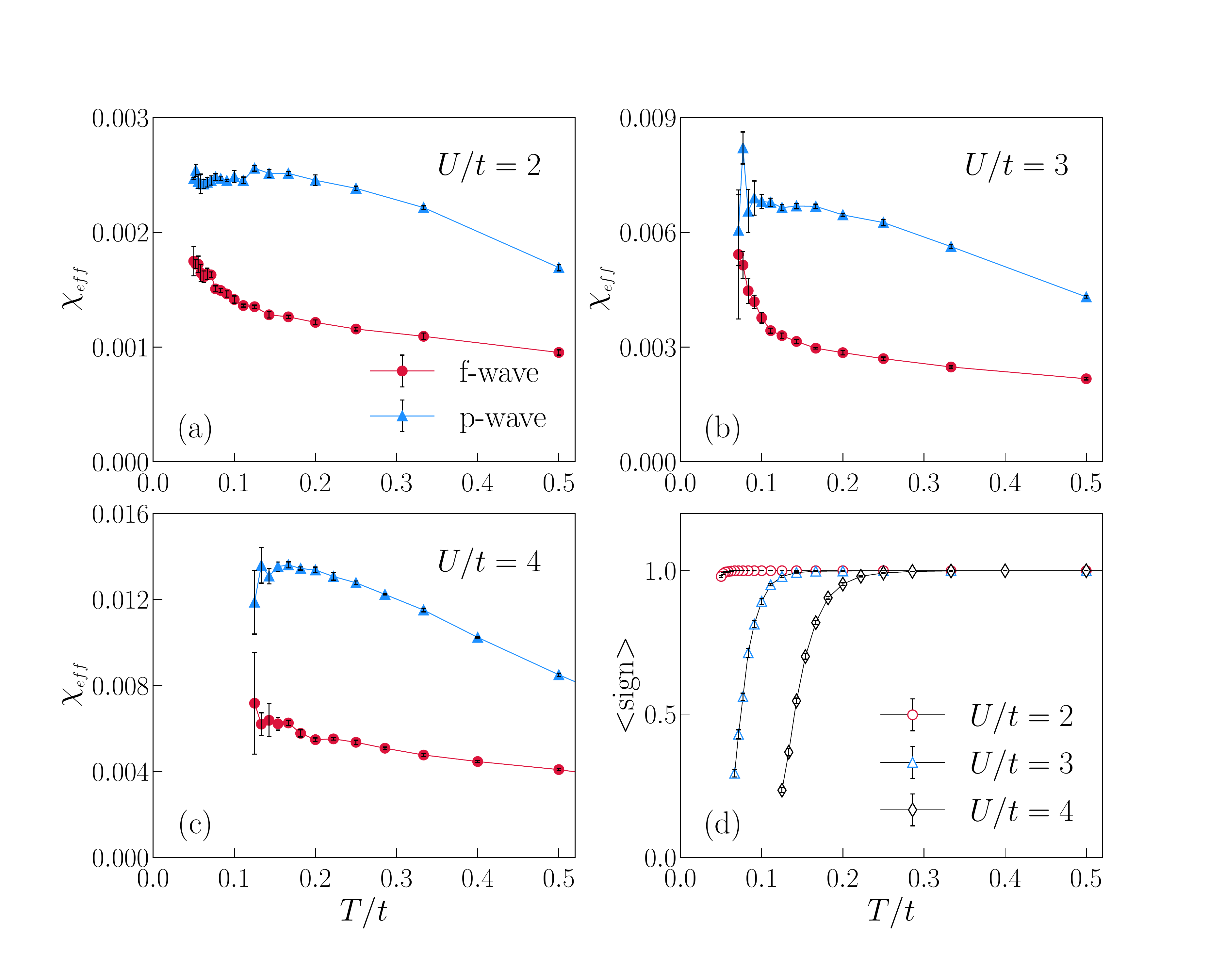} \caption{The effective susceptibility of the dominating triplet pairing
channels as a function of temperature for several values of $U$: (a) $U/t=2$, (b) $U/t=3$, and (c) $U/t=4$. (d) The average signs in (a)-(c) as a function of temperature. Here the system is at $5\%$ hole doping, corresponding to the average density $\rho=0.95$.}
\label{fig7}
\end{figure}

Figure \ref{fig7}(a)-(c) show the effective susceptibility of the $s_z=0$ spin-triplet pairings as a function of temperature for $U/t=2,3,4$ at $\rho=0.95$. As we move away from half filling into the doped region, the sign problem occurs and gets worse continuously with lowering the temperature, which limits the DQMC simulations to relatively high temperatures.
Although both $\chi_{eff}^{p}$ and $\chi_{eff}^{f}$ increase as the temperature is lowered in the high-temperature region, their low-temperature trends are different: the value of the $p$-wave pairing tends to saturate or decrease whereas the $f$-wave pairing continues to grow at the lowest temperature accessible to DQMC. Hence, if there is a possible superconducting instability, it would have a $f$-wave symmetry.
Besides, the value of $\chi_{eff}^{f}$ continuously increases with the interaction below $U/t=4$, implying the superconducting instability is gradually enhanced in this range of interaction strength.
In great contrast, $\chi_{eff}^{s^*,d}$ is negative over the temperature range simulated by DQMC, and decreases with increasing the interaction (see Appendix C), suggesting the singlet superconducting parings are not favored by the on-site Hubbard interaction. We also calculate the effective susceptibility of the $s_z = \pm 1$ pairings(see Appendix C). It is found the values of the singlet $s^*$- and $d$-wave parings are very close to zero, and the values for triplet $p$-wave and $f$-wave
ones are increasingly negative with decreasing temperature, suggesting these finite-$s_z$ paring channels are suppressed.

\begin{figure}[htbp]
\centering \includegraphics[width=8.5cm]{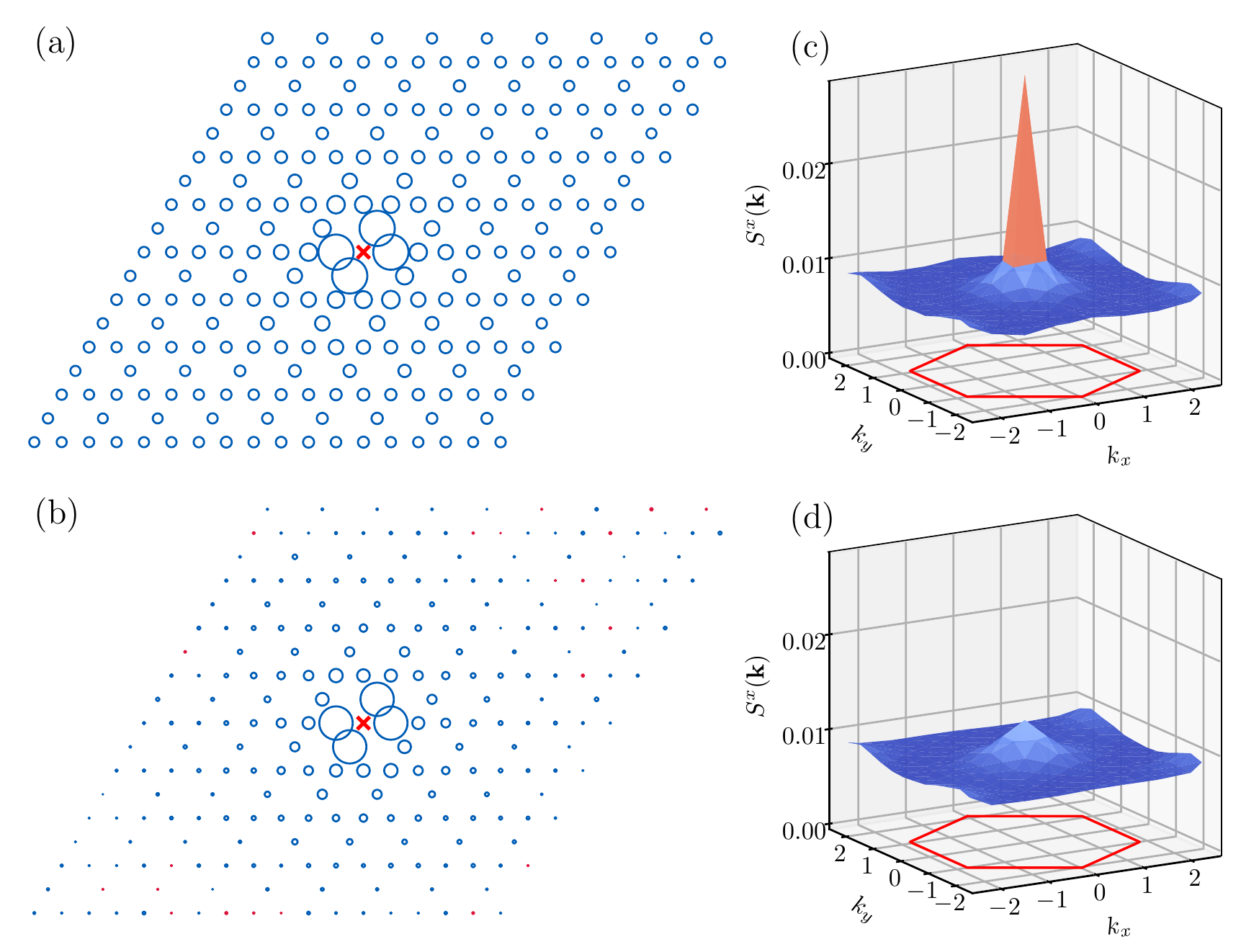} \caption{The transverse real-space spin correlations at $U/t=4$: (a) half filling ($\rho=1$); (b) $5\%$ hole doping ($\rho=0.95$). The magnitudes of the correlations are represented by
the radii of open circles. (c) and (d) are the corresponding structure factors $S^x({\bf k})$ of the spin correlations in (a) and (b), respectively.}
\label{fig8}
\end{figure}

To reveal the microscopic origin of the superconducting pairing interaction, the spin correlations are calculated for various values of $U$ at the densities $\rho=1$ (undoped) and $\rho=0.95$ (hole-doped). As the system is hole-doped, the values of the $x$ components of the spin correlations decrease quickly with the distance, and have tended to vanish at the scale of several lattice constants [see Fig.\ref{fig8}(a) and (b)]. The disappearance of the long-range FM order in the doped case is also reflected in the static structure factor. The sharp peak at ${\bf k}=(0,0)$ [see Fig.\ref{fig8}(c)], marking the existence of the FM order at half filling, is completely suppressed in the doped system [Fig.\ref{fig8}(d)].  These results imply that the FM order in the $xy$ plane is quickly destroyed by the doping, and the spin correlations become short-ranged, therefor generating strong transverse spin fluctuations. This is consistent with the enhancement of the triplet $s_z=0$ superconducting states, thus suggests the triplet superconductivity may be mediated by this $xy$-plane FM fluctuations. By contrast, the $z$-direction FM order remains and is almost unchanged after a small doping is introduced, which should account for the absence of the triplet $s_z=\pm 1$ pairings.

\section{Conclusions}

We have applied the mean-field theory and DQMC simulations to study the imbalanced kagome-lattice Hubbard model. Since the asymmetry of the kagome bands leads to the imbalance of the spin-up and -down electron densities, a $z$-direction FM order inherently exists in the system at half filling. After the interaction $U$ is turned on, the spin-$z$ FM order is first enhanced, and then decreases gradually after reaching its maximum strength. In the meantime, a $xy$-plane ferromagnetism develops above a critical interaction, which is estimated to be $U_c/t=3.65$ by finite-size scaling. The transverse magnetic transition is continuous, and is verified to belong to the 3D $XY$ universality class. We finally investigate the superconducting properties in the doped case. From the evolution of the pairing susceptibility as a function of temperature, we conclude that the triplet $f$-wave pairing will be dominant if there is a possible superconducting instability. Our these results are based on large-scale exact numerical simulations, and will deepen the understanding of the interplay between the electron-electron correlations and geometry frustrations.

\section*{Acknowledgments}
The authors thank Rubem Mondaini for carefully reading the manuscript
and for a number of suggestions for improving it. C.W and H.G. acknowledge support from the National Natural Science Foundation of China (NSFC) grant Nos.~11774019 and 12074022, the NSAF grant in NSFC with grant No. U1930402, the Fundamental Research
Funds for the Central Universities and the HPC resources
at Beihang University.
N.H. acknowledges support from NSFC Grants
No. 12022413, No. 11674331, the “Strategic Priority Research Program (B)” of the Chinese
Academy of Sciences, Grant No. XDB33030100, the ‘100 Talents Project’of the Chinese Academy of Sciences, the Collaborative Innovation Program of Hefei Science Center, CAS (Grants No. 2020HSC-CIP002), the CASHIPS Director's Fund (BJPY2019B03).
S.F. is supported by the National Key Research and
Development Program of China under Grant No. 2021YFA1401803,
and NSFC under Grant Nos. 11974051 and 11734002.

\appendix

\renewcommand{\thefigure}{A\arabic{figure}}

\setcounter{figure}{0}

\section{The imbalanced triangle-lattice Hubbard model}

\begin{figure}[htbp]
\centering \includegraphics[width=8.5cm]{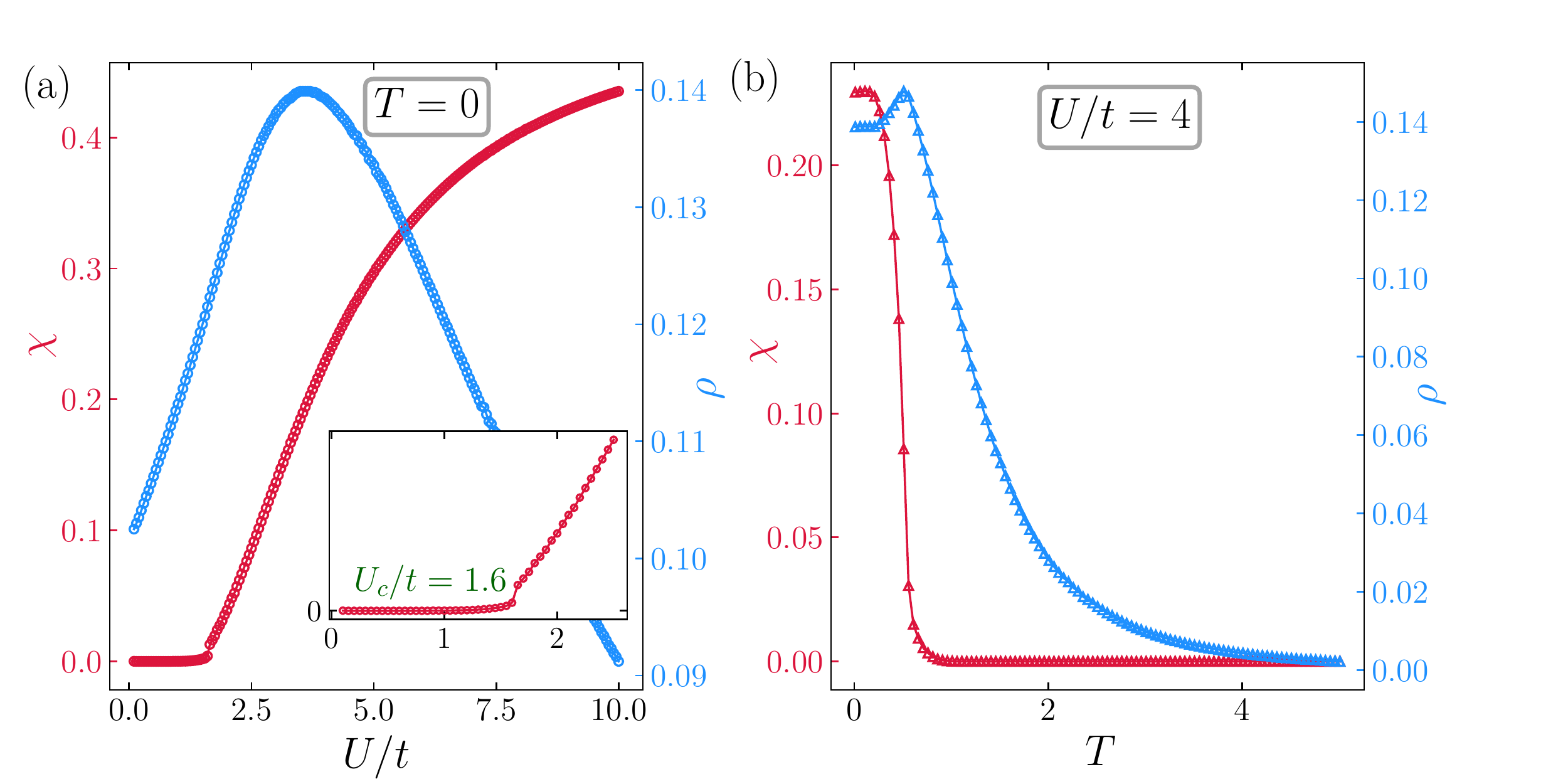} \caption{The order parameters $\chi,\rho$ in the imbalanced triangle-lattice Hubbard model: (a) as a function of $U$ at $T=0$; (b) as a function of $T$ at $U/t=4$. Inset in (a) shows enlarged view of the transition region of $\chi$, and the critical interaction is $U_c/t=1.6$, which is larger than that of the kagome geometry.}
\label{afig1}
\end{figure}

\begin{figure}[htbp]
\centering \includegraphics[width=8.5cm]{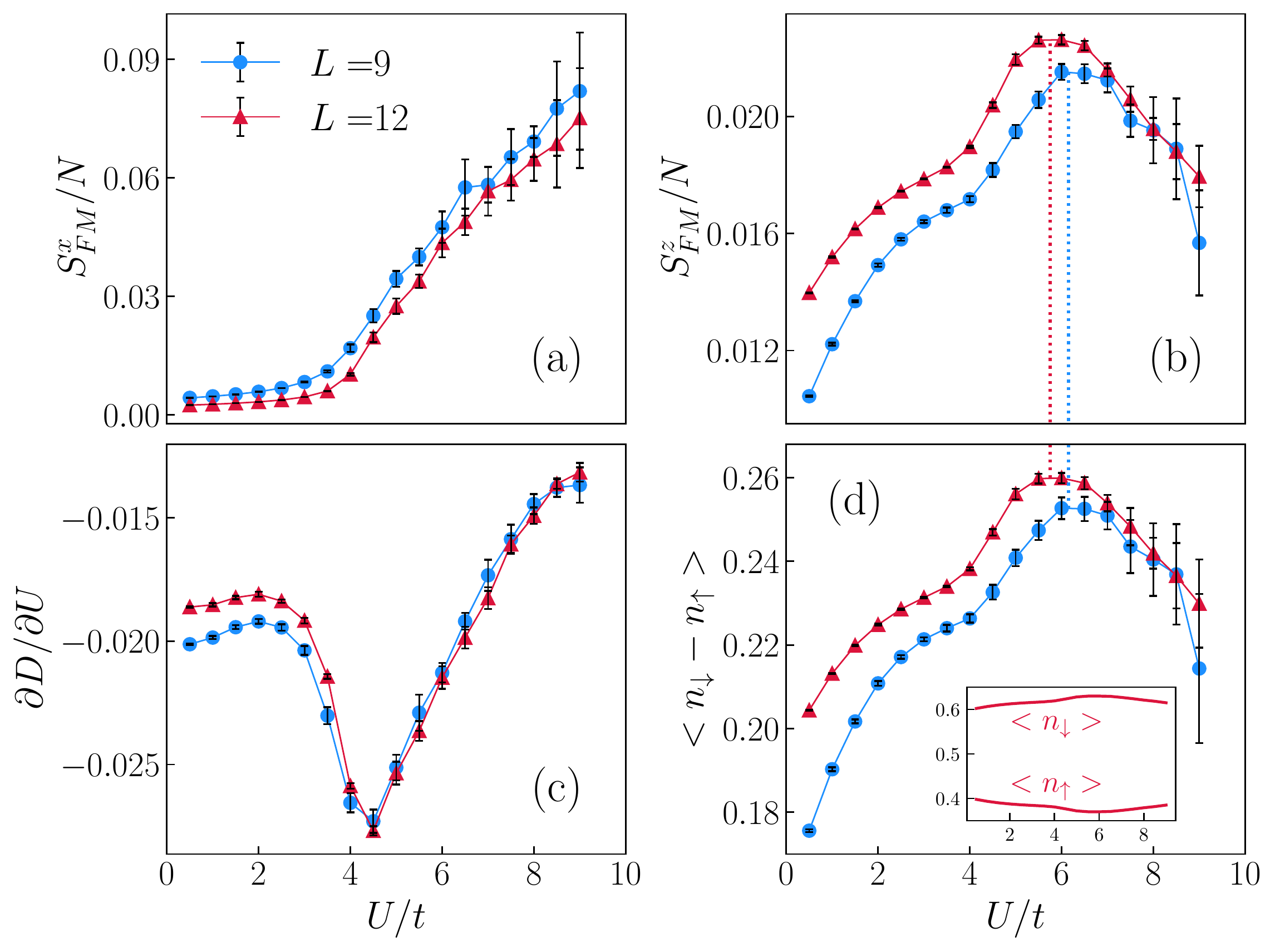} \caption{Several physical quantities as a function of $U$ for various $L$ at
$\beta t= 10$ in the imbalanced triangle-lattice Hubbard model: (a) $x$ component and (b) $z$ component of the FM structure factor; (c) the rate of change of the double occupancy; (d) local moment in the $z$ direction. Inset in (d) plots the up- and down-spin average densites for the lattice size $L=12$. The dashed lines in (b) and (d) mark the peaks of the curves, which coincide with each other at the same lattice size.}
\label{afig2}
\end{figure}

We also perform a mean-field analysis and DQMC simulations of the imbalanced triangle-lattice Hubbard model. As shown in Fig.\ref{afig1} and Fig.\ref{afig2}, the results are similar to those of the kagome-lattice system. The $z$-direction ferromagnetism, which is due to the imbalance of the densities of the different spin subsystems, is first enhanced by the Hubbard interaction. After reaching its maximum value, the corresponding order parameter $\rho$ begins to decrease monotonically, and tends to vanish at large $U$. The system develops the $xy$-plane ferromagnetical order above a critical value of the Hubbard interaction, which is estimated to be $U_c/t=1.6$ by the mean-field theory. Since the spin $SU(2)$ symmetry is broken by the modification of the hopping sign in the spin-down subsystem, both kinds of ferromagnetic states persist at finite temperatures. We also have applied DQMC method to calculate the $x$ component and $z$ component of the ferromagnetic structure factor, double occupancy, and local moment in the $z$ direction (see Fig.\ref{afig2}), and the numerical results are consistent with the mean-field predictions.

\section{The imbalanced square- and honeycom-lattice Hubbard models}

Square and honeycomb lattices are bipartite ones, and the non-interacting energy spectrum is symmetric with respect to the Fermi level at half filling. Unlike the frustrated systems, there is no aforemade ferromagnetic order here, and the on-site Hubbard interaction tends to induce AF correlations in the $z$ component of the spin. To qualitatively reveal the correlation physics of the imbalanced Hubbard models on the two geometries, we first perform a mean-field analysis.
Using the mean-field decoupling approach in Eq.(3), the interacting Hamiltonian becomes,
\begin{align}
H=&-t \sum_{\langle i j\rangle}(c_{i \uparrow}^{+} c_{j \uparrow}-c_{i \downarrow}^{+} c_{j \downarrow}+H . c .)+U \sum_{i} n_{i \uparrow} n_{i \downarrow} \\ \nonumber
\approx &-t \sum_{i_{A}, {\bf \delta_m}}\left( c_{i_{A}+{\bf \delta_m} \uparrow}-c_{i_{A} \downarrow}^{+} c_{i_{A}+{\bf \delta_m} \downarrow}+H . c . \right)\\ \nonumber
&+U \sum_{i_{A}}( n_{i_{A} \uparrow}\langle n_{i_{A} \downarrow}\rangle+\langle n_{i_{A} \uparrow}\rangle n_{i_{A} \downarrow}-\langle n_{i_{A} \uparrow}\rangle \langle n_{i_{A} \downarrow}\rangle \\ \nonumber
&-\left\langle S^{-}_{i_A}\right\rangle S^{+}_{i_A}-\left\langle S^{+}_{i_A}\right\rangle S^{-}_{i_A}+\left\langle S^{+}_{i_A}\right\rangle\left\langle S^{-}_{i_A}\right\rangle ) \\ \nonumber
&+U \sum_{i_{B}}( n_{i_{B} \uparrow}\langle n_{i_{B} \downarrow}\rangle+\langle n_{i_{B} \uparrow}\rangle n_{i_{B} \downarrow}-\langle n_{i_{B} \uparrow}\rangle \langle n_{i_{B} \downarrow}\rangle \\ \nonumber
&-\left\langle S^{-}_{i_B}\right\rangle S^{+}_{i_B}-\left\langle S^{+}_{i_B}\right\rangle S^{-}_{i_B}+\left\langle S^{+}_{i_B}\right\rangle\left\langle S^{-}_{i_B}\right\rangle ),
\end{align}
where $A, B$ denote the two sublattices (the square lattice also has an antiferromagnetic two-site unit cell), and ${\bf \delta_m}$($m=1,2,3,4$ for square lattice and $m=1,2,3$ for honeycomb lattice) represents the primitive vectors connecting the nearest-neighbor sites of any A-sublattice site. To incorporate the AF order, we write the averages of the operators as: $\langle n_{i_A,\uparrow(\downarrow)}\rangle=\frac{1}{2}\pm \rho, \langle n_{i_B,\uparrow(\downarrow)}\rangle=\frac{1}{2}\mp \rho, \langle S_i^{+}\rangle=\langle S_i^{-}\rangle=\chi$. Under a four-atomic basis $\psi_{\bf k}=(c_{A,{\bf k}}^{\uparrow},c_{B,{\bf k}}^{\uparrow},c_{A,{\bf k}}^{\downarrow},c_{B,{\bf k}}^{\downarrow})^{T}$, the Hamiltonian in the momentum space is,
\begin{align}
H_{MF}&=\sum_{\bf k}\psi_{\bf k}^{\dagger}{\cal H}({\bf k})\psi_{\bf k}+E_0,\\ \nonumber
{\cal H}({\bf k})&=\left(\begin{array}{cccc}
h^{-} & f({\bf k}) & -U\chi & 0  \\
f*({\bf k}) & h^{+} & 0 & -U\chi  \\
-U\chi & 0 & h^{+} & -f({\bf k}) \\
0 & -U\chi & -f*({\bf k}) & h^{-} \\
\end{array}\right),
\end{align}
where $h^{\pm}=\frac{1}{2}\pm \rho$, $f({\bf k})=-t\sum_{m}e^{i{\bf k}\cdot {\bf \delta}_m}$, and $E_0=2NU\chi^2+2NU\rho^2-NU/2$. At zero temperature, the order parameters $\rho, \chi$ are obtained by solving the self-consistent equations
\begin{align}
&\rho=-\frac{1}{4 N U} \frac{\partial E_{kin}}{\partial \rho} \\ \nonumber
&\chi=-\frac{1}{4 N U} \frac{\partial E_{kin}}{\partial \chi},
\end{align}
which is obtained by minimizing the total energy. Here $E_{kin}=\sum_{{\bf k},i\in occ}E_{\bf k}^{(i)}$ is the kinetic energy with $E_{\bf k}^{(i)}$ the eigenenergy of the $i$th band at the momentum ${\bf k}$, and the sum is over all negative eigenenergies, corresponding to the occupied states in the ground state.

\begin{figure}[htbp]
\centering \includegraphics[width=8.5cm]{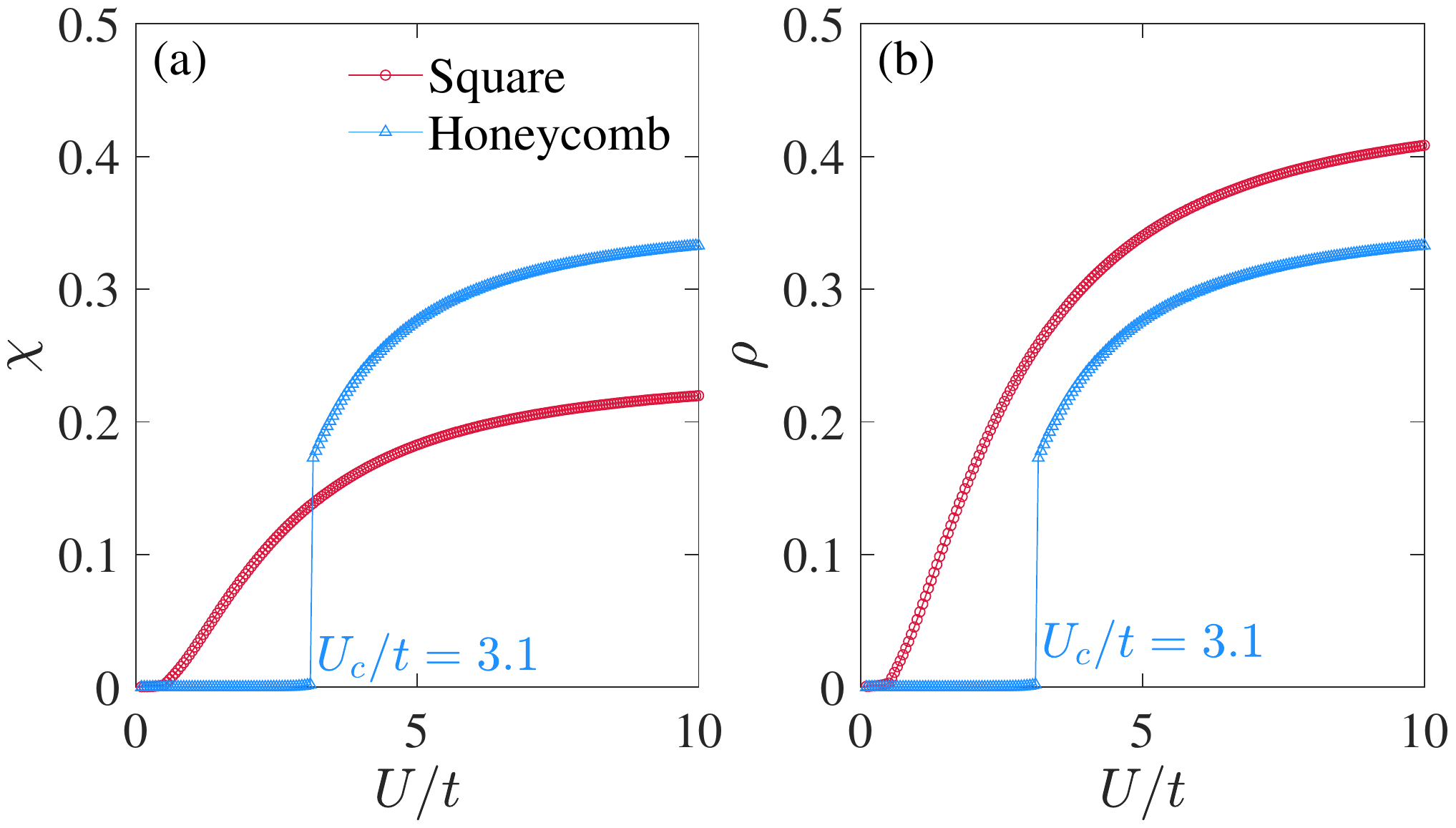} \caption{The order parameters of the imbalanced square- and honeycomb-lattice Hubbard models as a function of $U/t$ at $T=0$: (a) $\chi$ and (b) $\rho$. There is a critical interaction in the honeycomb geometry, which is $U_c/t=3.1$.}
\label{afig3}
\end{figure}

Figure \ref{afig3} plots $\rho,\chi$ as a function of $U$. Due to the vanishing density of states at the Fermi energy on honeycomb lattice, the magnetic transition occurs at a finite critical interaction $U_c/t=3.1$. Moreover, the critical values of the $z$-direction antiferromagnetism and $xy$-plane ferromagnetism are the same, suggesting they develop simultaneously. For the case of square lattice, $\rho, \chi$ continuously increase with $U$. It is well known that the long-range AF order exists for all $U>0$ in the normal square-lattice Hubbard model. Although there is a visible transition at small $U$, it may be due to finite-size effect, and it is expected that an infinitesimal $U$ can induce the magnetic order in the imbalanced square-lattice Hubbard model. For both geometries, $\rho,\chi$ saturate to a finite value at large $U$, which is in great contrast to the situation on frustrated lattices where the $z$-direction magnetism vanishes finally.

\begin{figure}[htbp]
\centering \includegraphics[width=8.5cm]{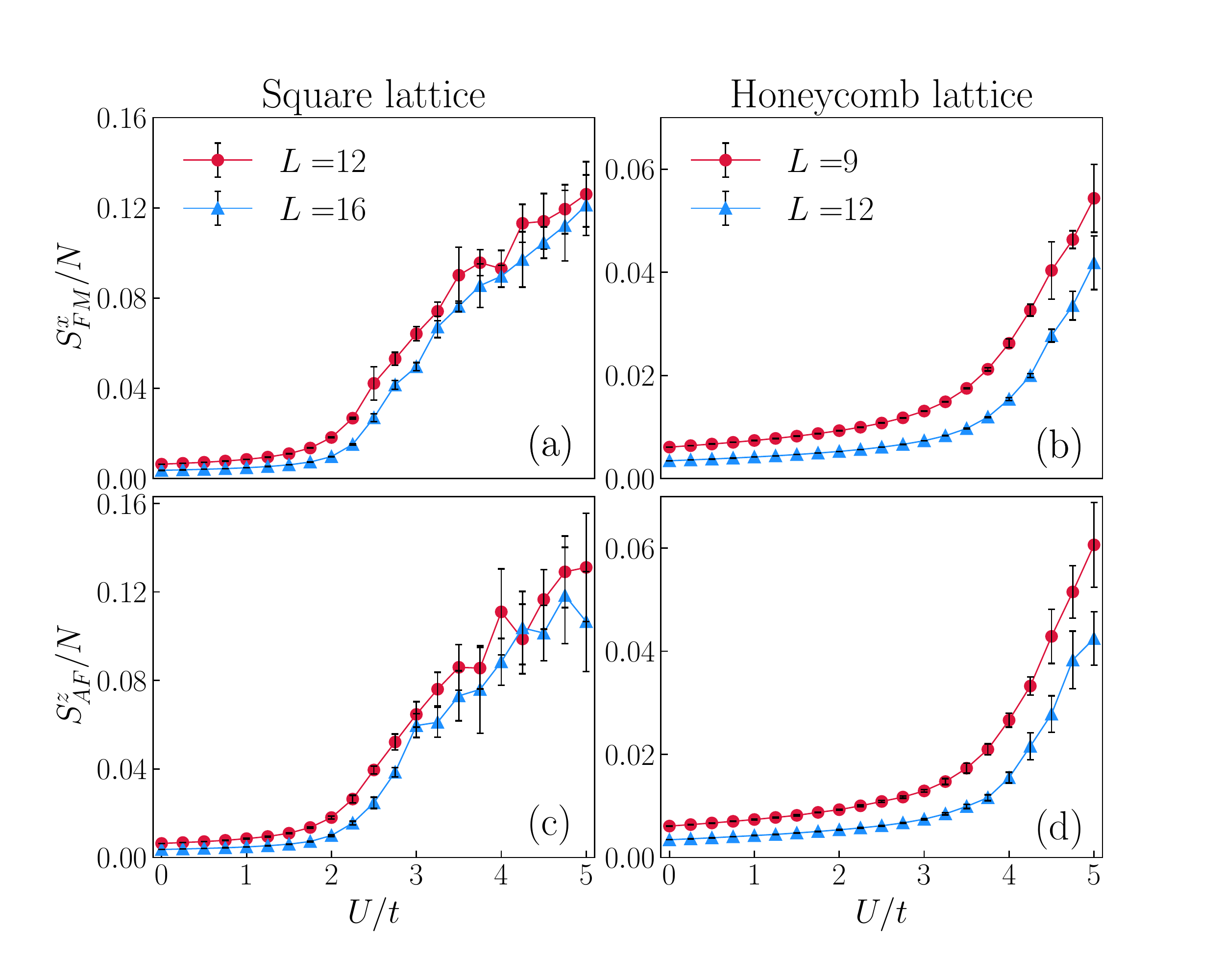} \caption{The $xy$-plane FM structure factor and the $z$-direction AF structure factor as a function of $U$ for: (a) and (c) square lattice; (b) and (d) honeycomb lattice. Here the DQMC simulations are performed at the inverse temperature $\beta t = 10$.}
\label{afig4}
\end{figure}

The imbalanced square- and honeycom-lattice Hubbard models are then simulated by DQMC. As shown in Fig.\ref{afig4}, both $S^{x}_{FM}$ and $S^{z}_{AF}$ increases continuously with the Hubbard interaction. While the values are negligibly small at weak couplings, they become large finite when the interaction is strong. This behavior implies a continuously magnetic transition occurs at a critical interaction, which is consistent with the mean-field theory. Compared to the case of the honeycomb geometry, the curves of the magnetic structure factors are shifted to the left a lot, suggesting the critical value of the square lattice is much smaller than that of the honeycomb one. It is also noted that the curves of $S^{x}_{FM}$ and $S^{z}_{AF}$ on the same lattice are almost identical. Hence the two kinds of magnetic phases are degenerate, and occur at the same critical interaction. After the critical value, the system can develop either the $xy$-plane FM or the $z$-direction AF long-range order in the ground state.

The situation can be better understood in the large-$U$ limit, when the low-energy physics is described by the spin Hamiltonian Eq.(9) in the main text. Subsequently, the spin model can be mapped onto the hardcore Bose-Hubbard model with nearest-neighbor repulsion, which has been widely investigated on square and honeycomb lattices. The model parameters in Eq.(9) correspond to the Heisenberg point of the phase diagram, locating on the boundary separating the charge-density-wave state and the superfluid. In the spin system, the counterparts of the above two bosonic phases are just the $z$-direction AF and $xy$-plane FM states, respectively.

\section{More DQMC results of the superconducting properties}

\begin{figure}[htbp]
\centering \includegraphics[width=8.5cm]{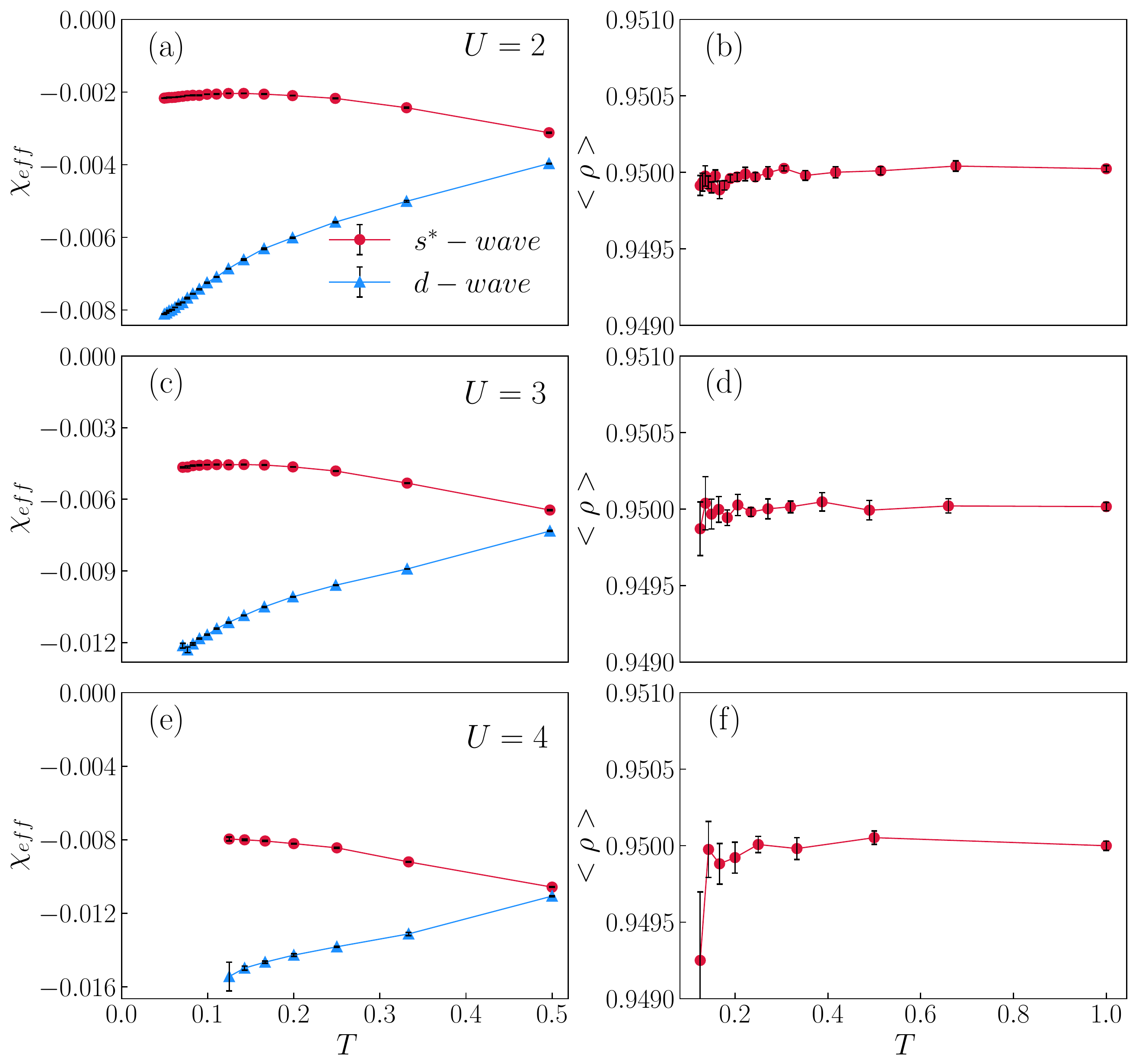} \caption{The effective pairing susceptibility of $s^{*}$- and $d$-wave
channels as a function of temperature for: (a) $U/t=2$; (c) $U/t=3$; (e) $U/t=4$. Here the system is at $5\%$ hole doping. (b), (d), and (f) show the average density at the manually determined chemical potential targeting the fixed density $\rho=0.95$ at the interaction strengths in (a), (c), and (e), respectively.}
\label{afig5}
\end{figure}

In the main text we have demonstrated the pairing susceptibility of the dominating spin-triplet pairing symmetry in the hole-doped case. Here we provide further justification, by investigating other symmetry channels, in complement to Fig.7 in the main text. Figure \ref{afig5} shows the effective susceptibility for $s^*$- and $d$-wave singlet pairings. $\chi_{eff}^{s^*,d}$ is negative over the temperature range simulated by DQMC, and decreases with increasing the interaction. Besides, the $d$-wave pairing is increasingly negative with decreasing temperature. These results suggest the above two singlet pairing symmetries are suppressed by the on-site Hubbard interaction.

\begin{figure}[htbp]
\centering \includegraphics[width=8.5cm]{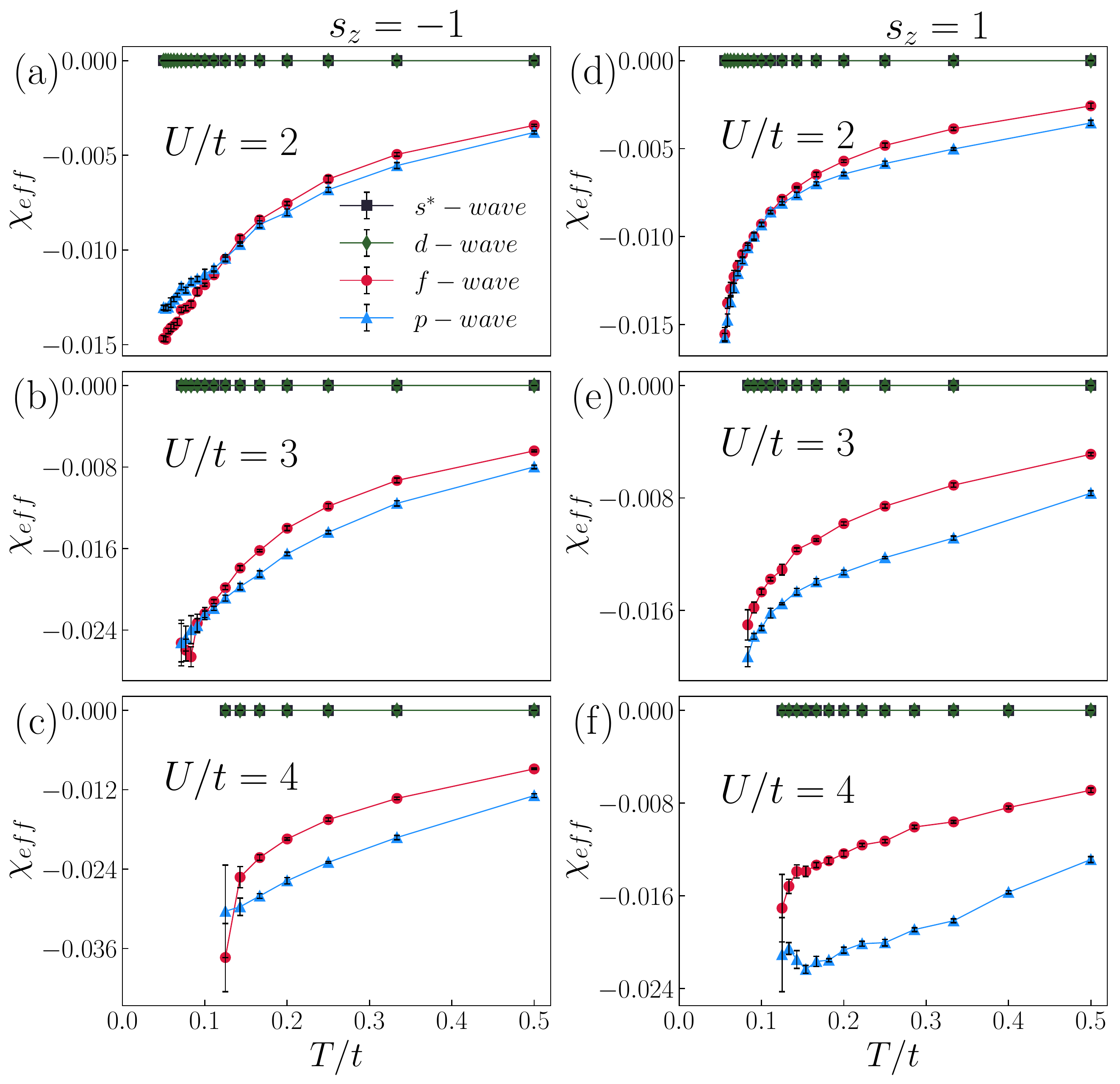} \caption{The effective pairing susceptibilities of various pairing
channels as a function of temperature for the $s_z=-1$ pairing operator: (a) $U/t=2$; (b) $U/t=3$; (c) $U/t=4$. Here the system is at $5\%$ hole doping. (d), (e), and (f) show the effective pairing susceptibilities for the $s_z=1$ pairing operator at the interaction strengths in (a), (b), and (c), respectively.}
\label{afig6}
\end{figure}

We also calculate the effective susceptibility of the $s_z = \pm 1$ pairings. As shown in Fig.\ref{afig6}, the values of the singlet $s^*$- and $d$-wave parings are very close to zero, and the values for triplet $p$-wave and $f$-wave
ones are increasingly negative with decreasing temperature, suggesting these finite-$s_z$ paring channels are suppressed.

\bibliography{ddirac}


\end{document}